# A Proposal for an Electron-Transfer Mechanism of Avian Magnetoreception

Shao-Qing Zhang[1,2,†]

[1] Department of Chemistry, University of Pennsylvania, Philadelphia, PA 19104-6396.
[2] Department of Pharmaceutical Chemistry and the Cardiovascular Research Institute, University of California at San Francisco, San Francisco, CA 94158-9001.
[†] zhangsh@sas.upenn.edu

**Abstract**

In spite of many years of research, the mechanism of avian magnetoreception remains a mystery due to its seemingly insurmountable intricacies. Recently Xie and colleagues proposed that IscA1 can act as a protein biocompass due to the measured intrinsic ferromagneticity, and thus named it MagR. However, Meister's calculations showed that the interaction energy of the magnetic moment of IscA1 with Earth's magnetic field is five magnitudes smaller than thermal fluctuation at room temperature. The other long-proposed compass protein is cryptochrome (Cry) with a mechanism of forming singlet-triplet radical pairs. However, this sensory mechanism still has no inferable information transmission routes. We propose a magnetoreception mechanism involving both the Cry and IscA1 proteins, through which photoinduced electrons are transported to redox-regulated ion channels to provoke neuronal responses. The structural features of the Cry-IscA1 complex that make it suitable for long-range electron transfer are discussed and how the magnetic effect leads to neuronal activity is described.

The mechanism of avian magnetoreception remains an enigma due to the interdisciplinary challenges from physics, chemistry, cell biology, neuroanatomy and animal behaviour [1]. The extreme weakness of Earth's magnetic field rules out almost all proteins from being the magnetoreceptor. Biogenic ferrimagnetic oxide magnetite, which is sensitive to magnetic fields, can be coupled to mechanosensitive ion channels for magnetoreception [2]. However, no evidence for its existence is found in the pigeon tissue cells that were implicated as magnetosensors [3]. Magnetotactic bacteria, which can generate magnetosomes, are the only known organism that utilizes biomineralized magnetites for direction sensing [4]. The molecular mechanism is still not fully elucidated and magnetosome formation involves a myriad of proteins which are not conservative in vertebrates [5].

A protein ferrimagnet biocompass model for magnetoreception was not proposed until Can Xie and colleagues found a Cry-binding protein IscA1 by screening. They claimed that this [2Fe-2S] cluster containing protein polymerizes and forms a ferrimagnet that is sensitive to geomagnetic field [6]. Meister calculated the magnetic moment of the Fe atoms in the IscA1 complex in the size measured in [6], and found that its interaction energy with geomagnetic field is five magnitudes smaller than $k_BT$ [7]. Winklhofer and Mouritsen pointed out that the spins of the Fe atoms in the [2Fe-2S] cluster in proteins are diamagnetic or paramagnetic, and thus in principle IscA1 or the Cry-IscA1 complex cannot be a ferrimagnet [8]. They also argued [8] that the claimed directional preference in the single-particle Cry-IscA1 complexes by electron microscopy cannot be deduced by the statistical data presented in [6]. Their calculation from the magnetization curve obtained for the Cry-IscA1 complex in solution [6] leads to a magnetic moment with geomagnetic interaction energy seven magnitudes smaller than thermal energy [8] (If the concentration of the Cry-IscA1 complex 3.8 mg/ml in solution [6] is considered, the energy is still about $10^5$ smaller than $k_BT$). Finally, there is an evident directional preference in the IscA1 crystals when applied by an external magnet [6]. We suspect that this is due to deposition of small iron oxide magnetite crystals, manifested by the tiny hairs, on the IscA1 crystals. As pointed by Meister [7], small magnetite crystals have strong magnetic moment due to high density of Fe atoms with strong exchange interaction, and can thus be very sensitive to external magnetic field. IscA1 has strong affinity with iron beads [6], which is the likely contamination source of magnetite crystals. Meanwhile, iron-sulphur clusters are generally unstable in aerobic environment, even with reducing regents [9]. The experiments for IscA1 should be performed in an anaerobic condition, as degraded iron-sulphur clusters can also contribute to iron contamination. Neither theory nor experiment can confer to IscA1 the identity as a ferrimagnet biocompass yet.

The other candidate of magnetoreceptor protein is Cry, which acts by interaction between geomagnetic field and the radical pairs produced by photoinduced electron transfer. This mechanism was first proposed by Klaus Schulten and colleagues [10], and later Cry was found to be the best protein candidate [11]. Photoexcitation generates a pairs of radicals inside Cry from the FAD cofactor and one tryptophan residue (W324) [12], and the radical pairs can form singlet and triplet states, which interconvert when a magnetic field is applied. The radical-pair model is the best studied magnetoreception mechanism [13], and there is a line of *in vivo* evidence implicating involvement of Cry in magnetosensation [14-25]. The radical-pair model explains the angular sensitivity in detection of external magnetic field by modulated weight in singlet/triplet yield, which represents different biochemical products. Computer simulations on coherent spin pairs are able to predict a high angular precision, less than 5°, for detection of the magnetic field direction [26]. However, the signalling pathway between a neurological signal and the singlet/triplet yield is lacking.

Based on the long lifetime of charge separation in Cry [27], angular sensitivity of radical pairs to magnetic field direction [26], the binding model of the Cry-IscA1 complex [6], and the neuronal activity obtained by Baines and colleagues [22], we propose another mechanism oriented by the neuronal signalling pathway of avian magnetoreception. The basis of this mechanism is that photoexcited electrons in Cry are transferred to the associated IscA1, and are propagated along the IscA1 polymer to a redox-active partner that activates ion channels. An external magnetic field affects the electron transfer process and thus changes the firing rate of action potentials so that magnetic neurological responses are produced. Photoinduced electron transfer in Cry is established by experiments [28], and a long lived charge separation [27] helps enhance electron transfer for further delivery. When the electrons are transmitted to the IscA1 polymer, they can hop between the [2F-2S] clusters, and finally reach a redox-active sensor as the electron acceptor. The sensors regulate the associated ion channels to effect neuronal activities [29, 30]. In other words, the neuronal signal modulation pattern reflects the direction of the magnetic field from the frequency change of electron transport along the IscA1 polymer.

Physical association between Cry and IscA1 is vital for detection of magnetic field direction. Magnetoreceptor molecules as an ensemble should have good alignment in their orientation, otherwise the sensing signals can obscure each other. In cellular environments, thermal fluctuations impact all the proteins and larger proteins or complexes have smaller mean displacements. Cry is a water soluble protein, and the thermal fluctuations can be influential even when binding to a membrane protein, because membrane proteins also rotate inside the lipids and lipid bilayers are also fluctuating [13]. The case is different as IscA1 can polymerize. When Cry binds to oligomeric IscA1, the latter is more rigidified and stronger polymerization is induced. Based on the structure model in [6], the ratio between Cry and IscA1 is 1:2. Longer IscA1 polymers can provide a larger and more stable binding surface for Cry binding. When long copolymers are formed, the influence of thermal fluctuations is minimal. The copolymers can line along the long shaft of the photoreceptor cells as shown in [31]. Therefore Cry-IscA1 copolymers have great directionality for sensing magnetic field directions. If Cry or IscA1 has a binding transmembrane protein, the effect is further strengthened. There is also a possibility that the copolymers can bundle during formation.

Polymeric Cry-IscA1 complexes also provide a conducive environment for long-range electron transfer. Blue light induces electron transfer to FAD in Cry [27, 32, 33], and the electron donor is a tryptophan or tyrosine residue on the protein surface [34, 35]. In the model by Xie, the C-terminal residues (525-539) of *Drosophila* Cry forms a helix, on the other site of the Trp triad with respect to the FAD cofactor, and binds to IscA on the helix (31-45) in the structure [6]. The C-terminal region of Cry is reported to be essential for magnetoreception *in vivo* [19, 22, 24]. Though there are no eukaryotic IscA1 structures, the bacterial polymeric IscA structure reveals that four IscA1 proteins with two pairs of adjacent [2Fe-2S] clusters at the core as a repeating subunit, spiral out to form a single-stranded helical supracomplex [6]. When Cry binds to the IscA1 polymeric complex, the former covers the latter also in a spiral fashion. At the Cry-IscA1 interface, in Cry there are several tryptophan, tyrosine and cysteine residues which are close to the FAD cofactors [6]. These amino acids are common participants in long-range electron transfer [36]. There are Tyr/Met (Tyr63/67, Met130) and Tyr (Tyr65/69/104) triads in *Drosophila* and *C. livia*, respectively, in IscA1. Two residues (Tyr67, Met130 in *Drosophila,* Tyr69/104 in *C. livia*) are close to the Cry-IscA1 interface, and the other one is very close to the [2Fe-2S] cluster, according to the predicted structures by the I-TASSER server [37-39]. The electron transfer route can be from FAD to some residues at Cry-IscA1 interface, and then to the [2F-2S] clusters, and then to the electron-transfer residues in Cry, and then to IscA1. In other words, photoinduced electrons hop between IscA1 and Cry in the long-range electron transfer passage to the final acceptor. Structure determination of the Cry-IscA1 complex, followed by experimental and computational work [40], is essential in elucidation of the transfer routes. There are three sorts of physical features that help the long-range electron transfer process. (1) The long lived charge separation facilitates electron transfer to IscA1 from FAD [27, 28], otherwise charge recombination happens before further delivery. (2) On top of photoreceptor cells for light exposure, Cry can be coherently photoexcited as they are found in a regular array on the IscA1 lattices. There are many examples of long-distance charge and energy transport in Nature due to coherent excitations [41, 42]. (3) As multiple copies of Cry can be excited, and there can be alternate electron transfer pathways within Cry and the tetramer IscA1 subunit, coherence in multiple tunnelling pathways can be induced to attenuate the decay of the electron transfer reaction [43, 44]. The Cry-IscA1 combinations with better electron transfer abilities, such as having more stable polymer formation and/or more efficient electron-transfer residues, could have evolved in migratory animals that depend on magnetoreception for direction. Before the IscA1 polymeric structure is solved, there is still a possibility that IscA1 polymers can transport electrons on their own.

Neural firing is the vital response of magnetoreception. Magnetic field has been shown to directly modulate Cry-mediated firing rate [22]. Holmes and colleagues shows that photoactivation of Cry is coupled to neural firing by the redox sensor Hyperkinetic (Hk) in the cytoplasmic auxiliary Kvβ subunits of the potassium ion channels [45]. They also observed that FAD reduction is required for neuronal light response, and that overexpression of SOD proteins or addition of $H_2O_2$ disrupts the cellular redox environment and thus abolishes neuronal response to blue light [45]. These findings are consistent with our proposal that FAD relays photoactivated electrons to redox sensors for regulating ion channel activities. A variety of ion channels are under redox regulation [29, 30], and the relative locations of the redox sensors are diverse [30]. In our scenario, either Cry or IscA1 is likely to bind to the redox sensors to transport the electrons to the latter. The redox sensors mediate the interaction between Cry/IscA1 and ion channels. High throughput screening could soon uncover the binding partners of Cry and IscA1. The distribution of Cry and IscA1 is broad in many cell types [46], but the sensors and the associated ion channels can be highly expressed only in specific neuronal cells. The sensor could be a component of the ion channels, or a protein complex consisting of one liaison protein and the ion channel component. Specificity of the action of the redox sensors should also be investigated [45]. Hk should be one of the redox sensor candidates for magnetosensation. Thus it is immediately interesting to test whether it is the case in *Drosophila*, and whether Hk is used for magnetoreception in *C. livia* and other migratory birds. Finally, it is worthwhile to mention that the Zhang and Lu groups have recently studied magnetic field induced neuronal responses conferred by IscA1 [47, 48]. Their results on calcium channel activities are contradictory. However, if the Lu group could investigate the firing of action potentials on the neuronal cells, more light can be shed on the role of IscA1 in magnetoreception, as ion channels permeating other kinds of ions could take part.

Detection of the direction of the magnetic field is achieved by the induced change in the neural firing rate. The firing rate is determined by the frequency of occurrences of electron transfer. The rate-limiting step is photoinduced electron transfer from Cry to IscA1. The life time of charge separation in Cry decides the likelihood of electron delivery to IscA1. The long life of charge separation [27] allows electron transfer from Cry to IscA1 to induce neuronal signal [22]. The application of magnetic field enhances neuronal activity [22] for the following reason. In the photoinduced electron transfer process, one intermediate step is formation of radical pairs. If the radical pairs are long lived, the direction of the magnetic field can be detected at a certain angle with pronounced conversion of triplets [13], which forbids charge recombination. Thus the magnetic field delays charge recombination and boosts the occurrences of electron transfer away from Cry. The narrow angular sensitivity range of the radical pairs to magnetic field direction [26] is conveyed to a sharp change in neural firing rate, which can be easily detected by migratory animals. Additionally, based on the results on $Cry^{W324F}$ and the C-terminal truncated Cry [22], we can see that the charge transfer from Cry reflects the interaction between Cry and IscA1 on the structure level [6]. Without a magnetic field blue light does not induce neuronal response through $Cry^{W324F}$, but an added magnetic field does [22]. This is because the magnetic field interacts the radical pairs excited by blue light. The lifetime of charge separation is so short in $Cry^{W324F}$ that it cannot afford electron transfer from Cry to IscA1. However, the applied magnetic field enhances the triplet yield to extend the lifetime of charge separation sufficiently for electron transfer to happen. The radical pairs involved are not yet included in the current radical-pair model [13]. The C-terminal (521-540) truncated Cry is constitutively active, and is not sensitive to external magnetic field [22]. From the structure model, the residues 498-518 and the C-terminus of Cry as two helices interact with IscA1 [6]. When the C-terminus is removed, the distance between FAD and the [2Fe-2S] cluster is small, and the energy cost for electron transfer from Cry to IscA1 becomes so low that the enhancement from magnetic field for charge separation is insensitive. Therefore the C-terminus of Cry is the negatively regulatory component for magnetoreception: the excited electron from Cry to IscA1 has to go through an energy barrier imposed by the C-terminus of Cry to induce signalling. How the timescales in the rates of electron transfer and neural firing will be an interesting problem to investigate. The magnetic-field dependent long range electron transfer here is a nonequilibrium process, and thus there is no $k_B T$-problem related to it [13, 49]. Above all, the influence of the magnetic field is to restrain charge recombination, and thus to increase the firing rate as shown in [22]. Magnetoreception is achieved by migratory birds by following the direction with a sharp change in neuronal response. The information processing of neural firing [50] is out of the scope of this proposal.

The mechanism of avian magnetoreception as an integrative problem of quantum biology [51] has intrigued scientists for many years. As the first proposal to directly connect magnetic response and neuronal activity, the hypothesis described here can hopefully inspire researchers from different disciplines in diverse aspects to further understand magnetic sensing in animals. We believe that magnetogenetics as a new field will be established when all components of magnetoreception are discovered, characterized, optimized, and applied for different scientific, engineering and biomedical purposes.

**Acknowledgments**
S.-Q.Z. is grateful to Dr. Jason Donald for reading and editing the manuscript. S.-Q.Z. was financially supported by the MRSEC program of NIH through a grant to the LRSM at the University of Pennsylvania.